\documentclass[preprintnumbers,aps,twocolumn,floatfix,superscriptaddress]{revtex4-1}
\usepackage{eso-pic,calc}
\usepackage{graphicx}
\usepackage{amsmath, amsthm, amssymb}
\usepackage{epsfig}
\usepackage{bm}
\usepackage{color}
\usepackage[colorlinks,linktocpage,linkcolor=black]{hyperref}

\def\beqr{\begin{eqnarray}}
\def\eqnr{\end{eqnarray}}
\def\beq{\begin{equation}}
\def\bc{\begin{center}}
\def\ec{\end{center}}
\def\eqn{\end{equation}}

\def\etl{$et ~al.$}

\begin{document}

\title{Critical neuronal avalanches in levels model under noisy drive}

\author{Abdul Quadir}
\affiliation{Department of Physics, Aligarh Muslim University, Aligarh 202 002, India}

\author{Haider Hasan Jafri}
\affiliation{Department of Physics, Aligarh Muslim University, Aligarh 202 002, India}

\author{Avinash Chand Yadav\footnote{jnu.avinash@gmail.com}}
\affiliation{Department of Physics, Institute of Science, Banaras Hindu University, Varanasi 221 005, India}

\begin{abstract}
{We consider a neuronal levels model that exhibits critical avalanches satisfying power-law distribution. The model has recently explained a change in the scaling exponent from 3/2 to 5/4, accounting for a change in the drive condition from no input to moderate strength, along with a relaxed separation of time-scale between drive and dissipation. To understand the robustness of the scaling features, we examine the effect of different noisy stimuli in the moderate input regime. Our tool of analysis is the scaling method. We compute scaling functions associated with the avalanche size distribution, revealing striking finite-size scaling. For a class of noisy drives, we find that the scaling exponent can take a value different from 5/4, with an explicit system size dependence of the distribution.   } 
\end{abstract}

\maketitle

\section{Introduction}
The emergence of spontaneous neuronal activity-- the so-called ``neuronal avalanches''-- has been a topic of continuing interest in computational neuroscience and statistical physics. Since the early work by Beggs and Plenz~\cite{Beggs_2003, Beggs_2004}, many experimental studies, performed both in vivo and in vitro systems, suggest that such a neuronal activity is a generic behavior. Besides the availability of high-resolution data~\cite{Friedman_2012}, it is also possible to record it at a single neuron level~\cite{Gal_2013}. Considerable efforts, made so far experimentally and theoretically, have attempted to uncover its extent and functional significance~\cite{Chialvo_2010, Kinouchi_2013, Brochini_2016, Stoop_2016, Levina_2017, Davidsen_2021}.

From the structural and functional aspects, one can consider the neuronal system a living example of a complex system that arises from many interconnected neurons (nodes and links). The nodes may fire (get activated) when they receive stimuli (either externally or internally from neighbors). While firing, the node transfers signals to its neighbors and goes into a refractory state. The linked nodes may fire and eventually generate a burst of activity. If we record the number of active nodes as a function of time with a suitable time resolution, there would be many pauses or frame(s) with no activity. One can refer to the event as an avalanche activity, with its size defined as the sum of active nodes of frames lying between two subsequent quiet frames. Strikingly, the probability distribution of the avalanche sizes decays in a power-law manner. The existence of power-law in nature is a sign of complexity, reflecting the observable lacks a characteristic scale. It is interesting to note that even diverse systems may have the same scaling exponents, and we can group them into a universality class.

Although a significant amount of work attempts to understand the origin of scaling in the neuronal system, it remains an active research topic. The observation of power law might have the following implications. Possibly, the neuronal system operates near a critical point of a continuous phase transition between active (super-critical) and absorbing (sub-critical) phases. The criticality remains functionally beneficial as noted by observations like optimal dynamic range~\cite{Kinouchi_2006}, information processing, and storage capacity~\cite{Haldeman_2005}. However, the problem arises if one asks how the system maintains itself at the critical point. Tuning of an external control parameter in such systems seems unlikely. Several studies~\cite{Bornholdt_2000, Bornholdt_2003, Herrmann_2006, Anna_2009} have examined the issue by incorporating the notion of evolving links (dynamical synapses). It suggests the plausible mechanism can be self-organized criticality (SOC)~\cite{Bak_1987, Bak_1996, Millman_2010, Hergarten_2012}. The SOC provides a simple way by which the system can spontaneously organize itself into a dynamical-critical state, where the avalanches exhibit scaling features. Also, a recent study suggests that a neutral theory can lead to the emergence of scaling features~\cite{Martinello_2017}.

Recent works have examined several fascinating aspects associated with neuronal avalanches, including the interplay of the dynamics of both node and topology~\cite{Manchanda_2013} and the role played by the population of inhibitory nodes~\cite{Larremore_2014}. Wang \etl~ suggest that stochastic oscillation (a feature with a typical characteristic scale) and SOC (without characteristic scale) can emerge in the same model by varying the external drive strength~\cite{Wang _2016}. The neural system dynamically grows, while self-organization keeps the system in the critical state~\cite{Yaroslav _2018}.

More recently, Das and Levina~\cite{Das _2019} show the pertinent observation of avalanche-like events in neuronal models with discrete states (cellular automaton). Most remarkably, the power-law exponent for the avalanche size distribution changes from a typical value of 1.5 to 1.25 when they change the external drive regime from no external input to moderate strength. Thus, the universality class of the system also alters. The exponent 1.25 is close to an experimentally reported value of 1.3~\cite{Yu_2017} obtained by using a corrected avalanche detection algorithm for a task performing system with an increased drive.

In such critical models, typically, the system is slowly perturbed and allowed to dissipate instantly until the system achieves a stable configuration. Thus, a clear separation of time scales exists between drive and dissipation, leading to avoid interaction among avalanches. In the moderate regime, a delivery of input occurs during the avalanche event. It captures a physically relevant feature: The neuronal system undergoes external stimulation even during the spontaneous burst, and such a condition is inevitable most times. A partly relaxation of the separation of time-scale can preserve criticality, since the avalanches are well separated.

Das and Levina~\cite{Das _2019} have examined two models to understand the moderate drive effect, with the relaxed separation of timescale: (i) Branching model and (ii) levels model. With rigorous analytical results and simulations, they have shown that both models hold the same behavior. Although the branching model is the standard model typically used to study neuronal activity, this is not analytically tractable in the moderate drive regime.
The neural levels model is solvable for the avalanches size distribution in both drive regimes. And it also captures minimal features of the physical system. It makes the latter model more attractive. Since the levels model offers a better analytical insight, we consider the levels model as our elementary framework for further investigation.

In this paper, we mainly examine the following issue. To what extent the scaling exponent is robust under different noisy drives with moderate strength. In this regime, the avalanche size has two contributions, initial size without input and secondary size generated because of the moderate drive. While the previous studies suggest that the exponent $\tau_s = 5/4$ is robust, we here show that the distribution of secondary avalanche size dominates. For a class of inputs, the exponent characterizing the secondary avalanche size can take a value different from 5/4, along with an explicit system size dependence in the distribution. Interestingly,  the prefactor of the distribution varies as $N^{-\theta}$ because of the finite-system size effect. Strikingly, the sum of two exponent $\tau_s +\theta$ remains $1.25 \pm 0.10$.

The plan of the paper is as follows. In Sec.~\ref{sec_2}, we present the model definition. Section~\ref{sec_3} offers an application of the scaling method to provide a framework for the scaling function, describing avalanche properties. In Sec.~\ref{sec_4}, we show the results for both no external and moderate input regimes. Finally, we summarize our results in Sec.~\ref{sec_5}.

\begin{figure*}[b]
	\centering
	\scalebox{0.68}{\includegraphics{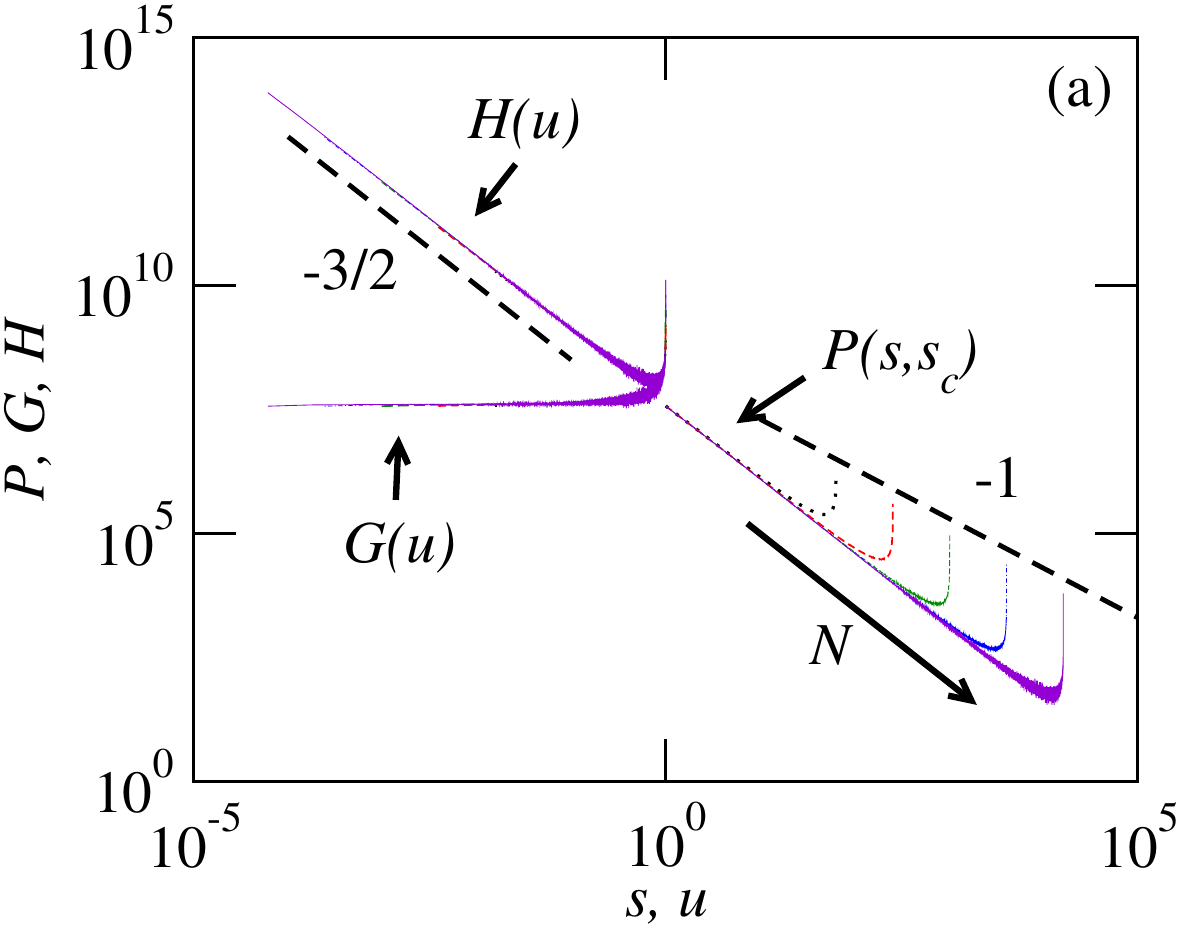}}
	\scalebox{0.68}{\includegraphics{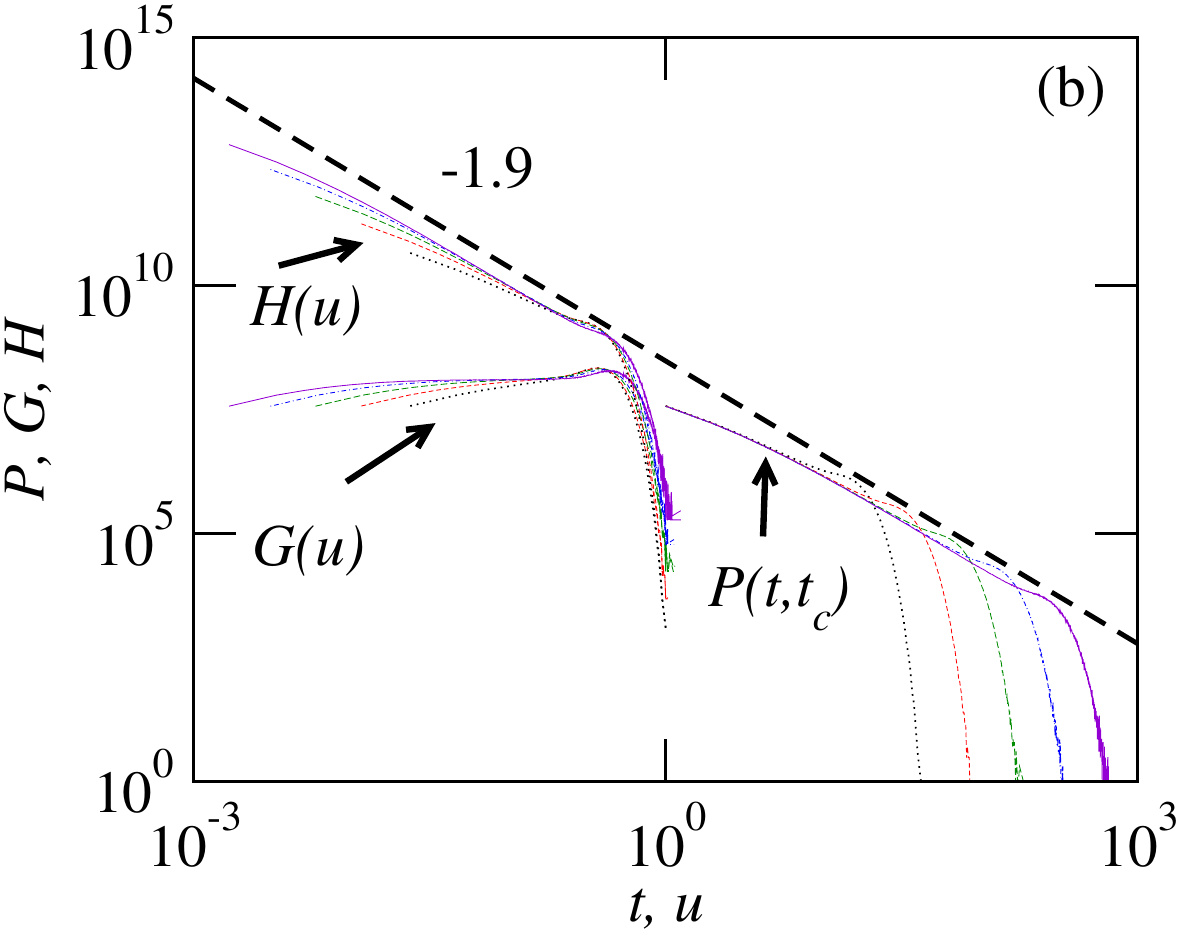}}
	\scalebox{0.68}{\includegraphics{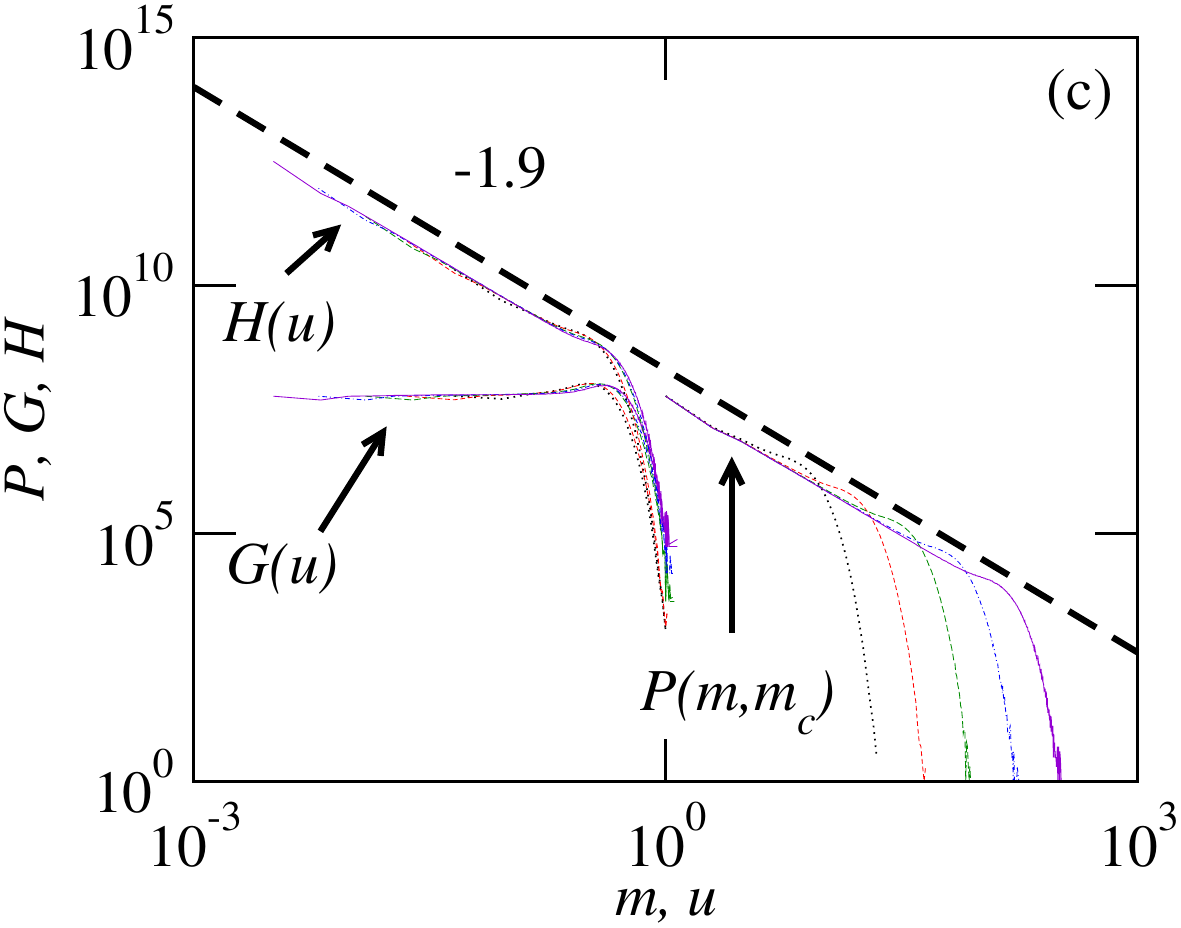}}
	\scalebox{0.66}{\includegraphics{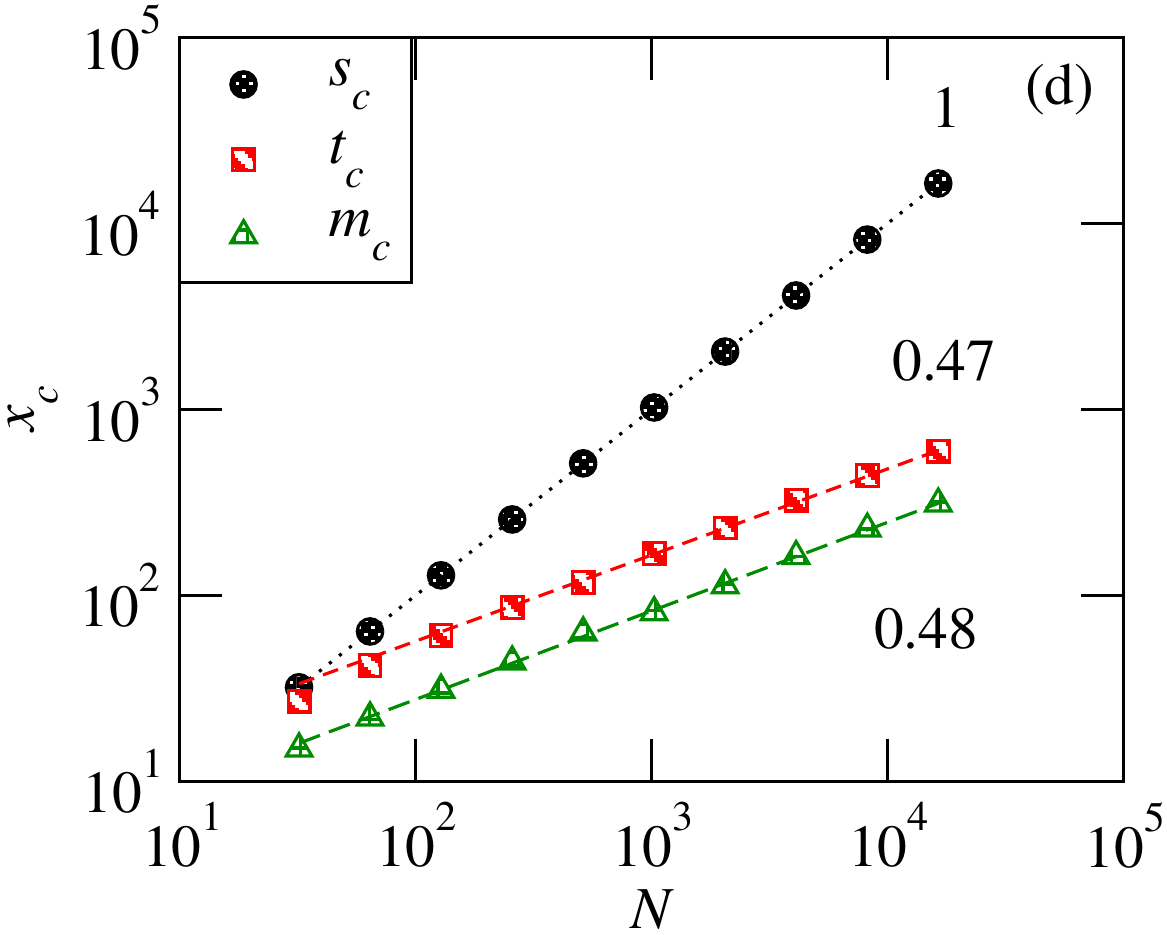}}
	\caption{No external input regime: (a) to (c) show the probability distribution function (PDF) $P(x,x_c)$ and the scaling functions  $G(u)$ or $H(u)$ for size $s$, duration $t$ and magnitude $m$. The total number of avalanches are $10^8$ and $N$ varies from $2^6, 2^8, \cdots$ to $2^{14}$.  The cutoff as a function of the system size is shown in (d).}
	\label{fig1} 
\end{figure*}

\begin{figure*}[t]
	\centering
	\scalebox{0.68}{\includegraphics{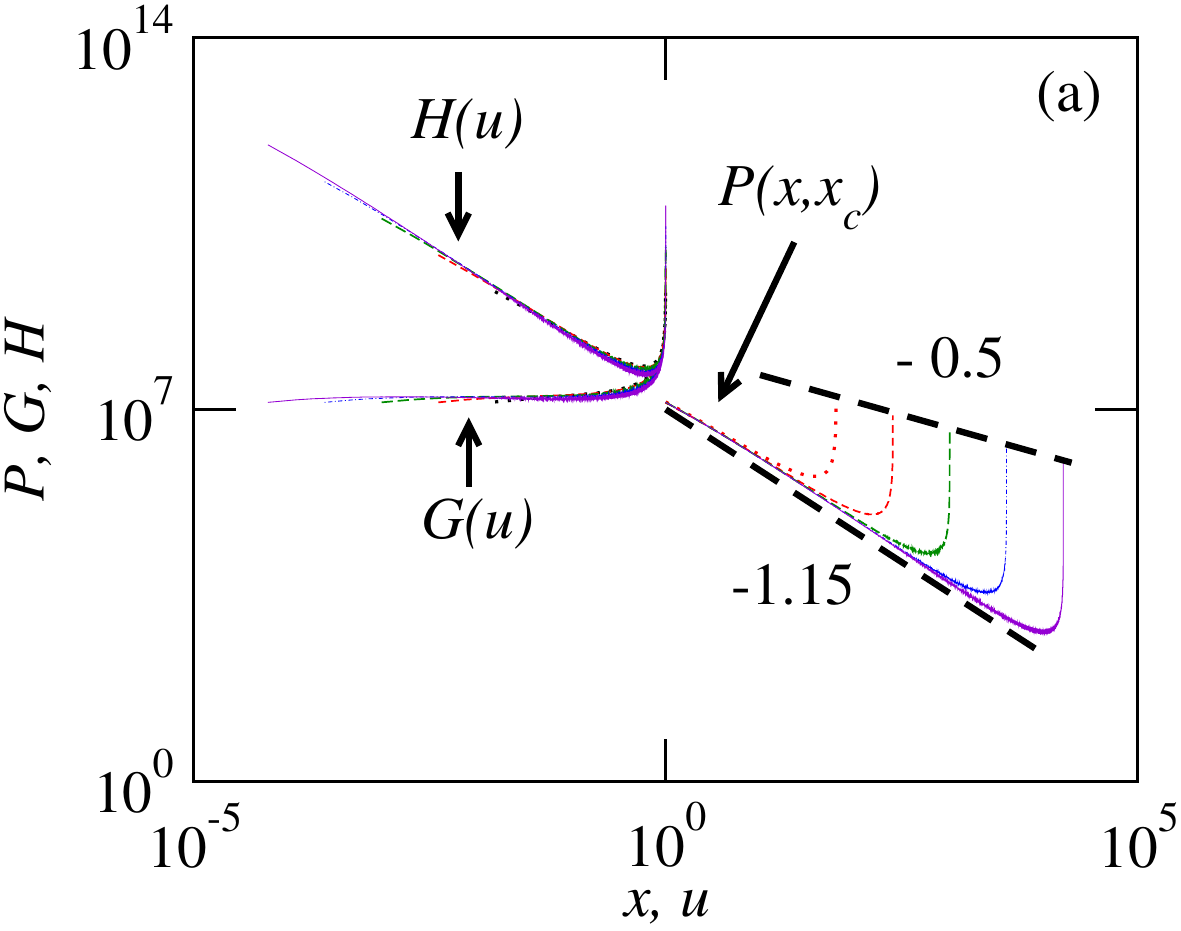}}
	\scalebox{0.68}{\includegraphics{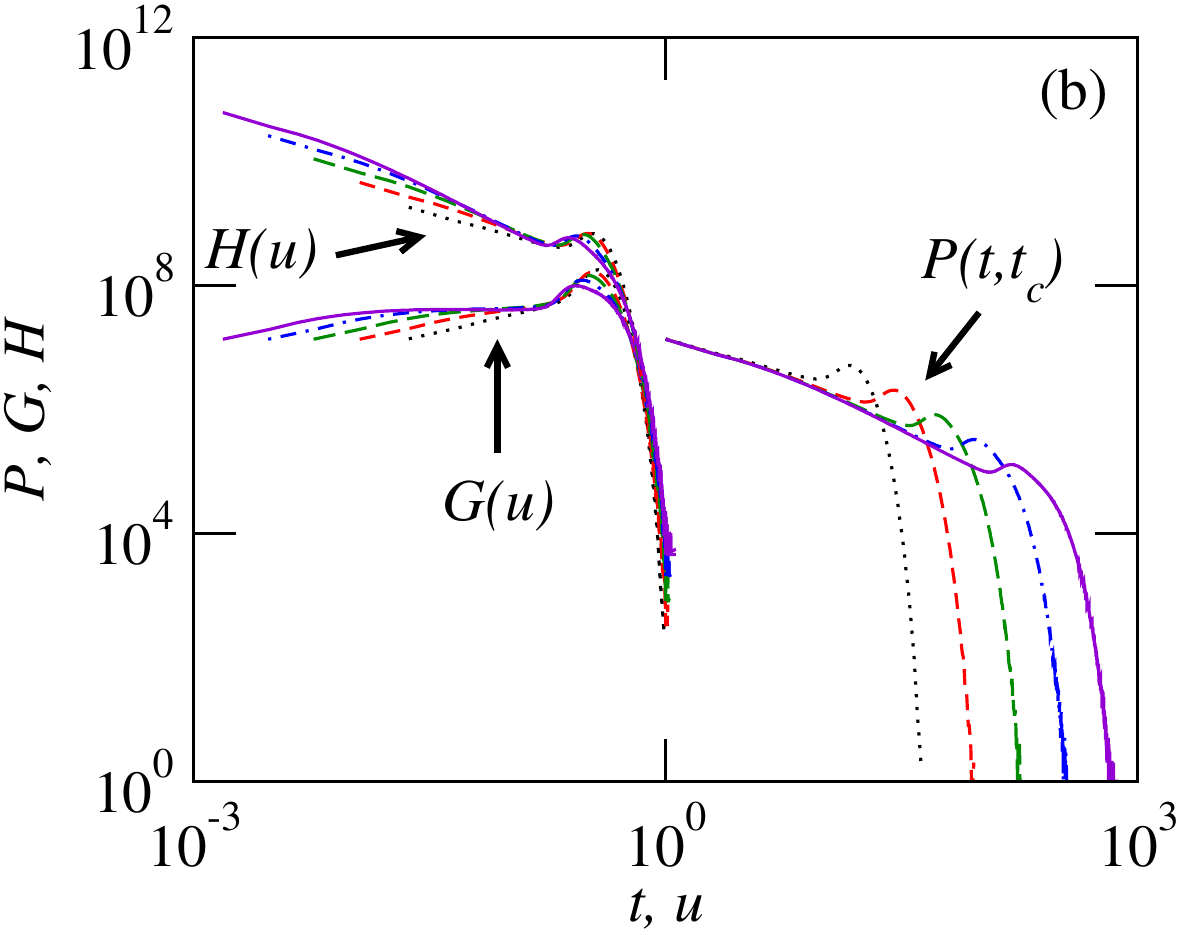}}
	\caption{ Same as Fig.~\ref{fig1}~(a) and (b), but in the moderate drive input regime. The size correspondence to the total size $x \to s+o$.}
	\label{fig2} 
\end{figure*}

\begin{figure}[t]
	\centering
	\scalebox{0.68}{\includegraphics{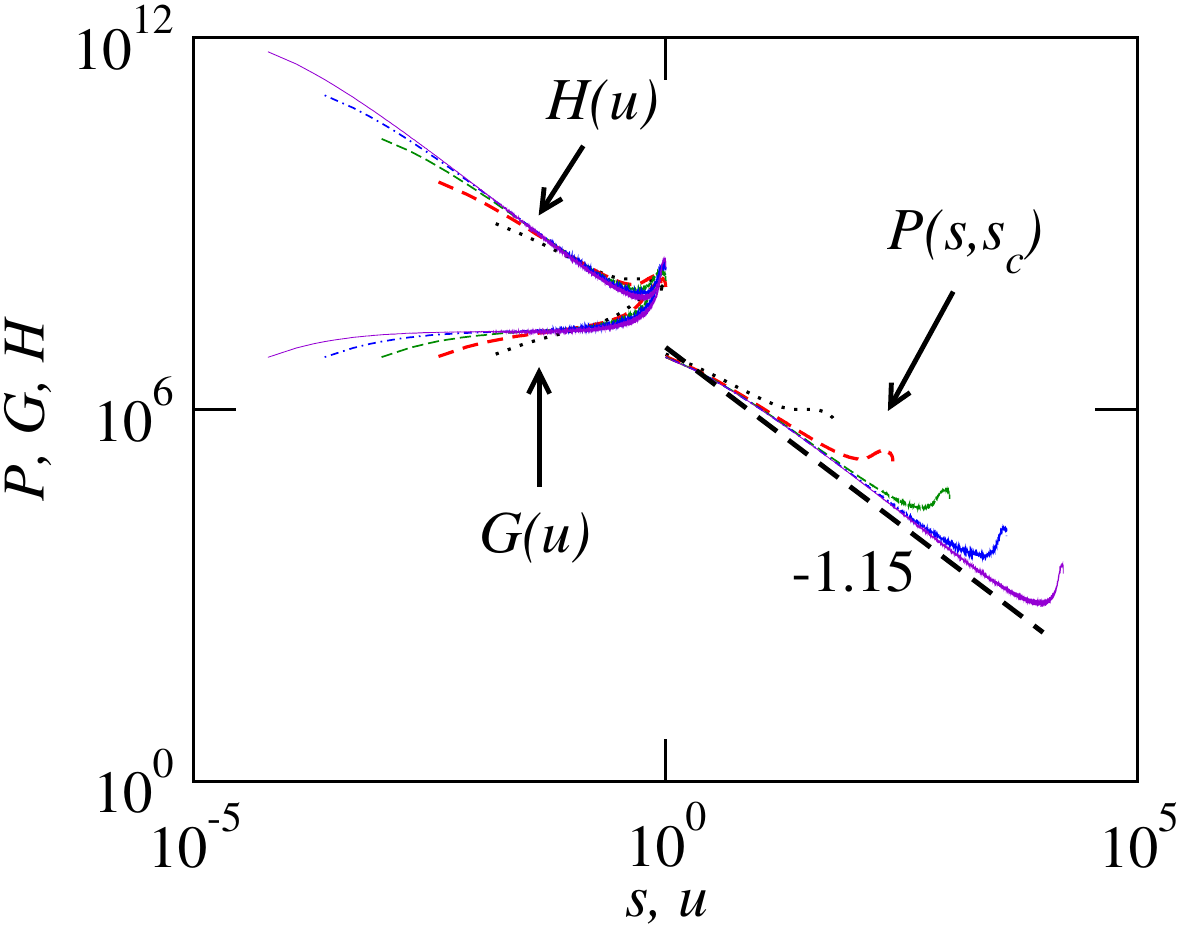}}
	\caption{ Same as Fig.~\ref{fig2}~(a), but for the secondary avalanche size.}
	\label{fig3} 
\end{figure}

\section{Model Definition}\label{sec_2}
The neuronal levels model is one of the simplest frameworks for describing neuronal activity. The system dynamics is a cellular automaton. The model comprises essential features and explains the experimentally observed critical avalanches in neuronal systems. In the no external input regime, the avalanche size distribution follows $P(s) \sim s^{-3/2}$ with a cutoff $s_c = N$. In the moderate external input regime, the exponent changes from 3/2 to 5/4.

The model definition is as follows. Consider a complete network with $N$ nodes (neurons), where each node connects with all remaining nodes, avoiding self-connection. Assign the membrane potential to each node as $E_i$, with a discrete value (level) between 1 to $M$. An active node $E_i = M$ fires by going into refractory state as $E_i \to 1$, and it transfers energy to the connected neighbors as $E_j \to E_j + 1$. This activity may trigger other nodes to fire, eventually forming a cascade event. We choose the threshold $M =N+1$, which corresponds to the critical case. The model dynamics is such that a node can fire only once during a single avalanche.

We initialize the system by assigning a random integer between 1 to $M-1$ to each node. For no external input regime, a randomly selected node fires and starts an avalanche. The update of the system occurs in parallel. When an avalanche event is over, a new independent event is generated with the same rules. 
The avalanche size (the number of firing nodes) measures a nonlinear response to such perturbations.

\section{Scaling functions}\label{sec_3}
The avalanche event is typically characterized by observable like $x \equiv $ size $s$, duration $t$, and maximum activity $m$, respectively. In general, the probability distribution for $x$ follows a decaying power-law~\cite{Yadav_2022_pre}
\begin{equation} 
P(x, x_c) = \begin{cases} Ax^{-\tau_x}x_c^{-\theta}, ~~~~~~~~~~~~~~~~{\rm for}~~ x\ll x_c,\\ {\rm rapid ~ decay~or ~growth},~ {\rm for }~ x\gg x_c, \end{cases} 
\end{equation}
where $x_c \sim N^{D_x}$, with $N$ being the number of nodes and $D_x$ is the cutoff exponent.  Here, a multiplicative factor $x_{c}^{-\theta}$ accounts for the finite system size dependence if the probability distribution function value decreases at a fixed $x$ with increasing the system size.

As the probability distribution is a homogeneous function of its arguments, we can write
\begin{equation} 
P(x, x_c) = \frac{1}{x^{\tau_x+\theta}} G(u) = \frac{1}{x_{c}^{\tau_x+\theta}}H(u),
\label{prob_2}
\end{equation}
where $u = x/x_c$. The scaling functions are given by
\begin{equation} 
G(u) \sim \begin{cases} u^{\theta}, ~~~~~~~~~~~~~~~~~~~~~~~~~~~~~~~{\rm for}~~ u\ll 1,\\ {\rm rapid ~ decay~or~ growth},~~~~~ {\rm for }~~u\gg 1, \end{cases} 
\end{equation}
and
\begin{equation} 
H(u) \sim \begin{cases} u^{-\tau_x}, ~~~~~~~~~~~~~~~~~~~~~~~~~~~~{\rm for}~~ u\ll 1,\\ {\rm rapid ~ decay~or~ growth},~~~~~ {\rm for }~~u\gg 1. \end{cases} 
\end{equation}
It is easy to note that the scaling functions isolate the two exponents describing characteristics of critical avalanches.

\begin{figure*}[t]
	\centering
	\scalebox{0.68}{\includegraphics{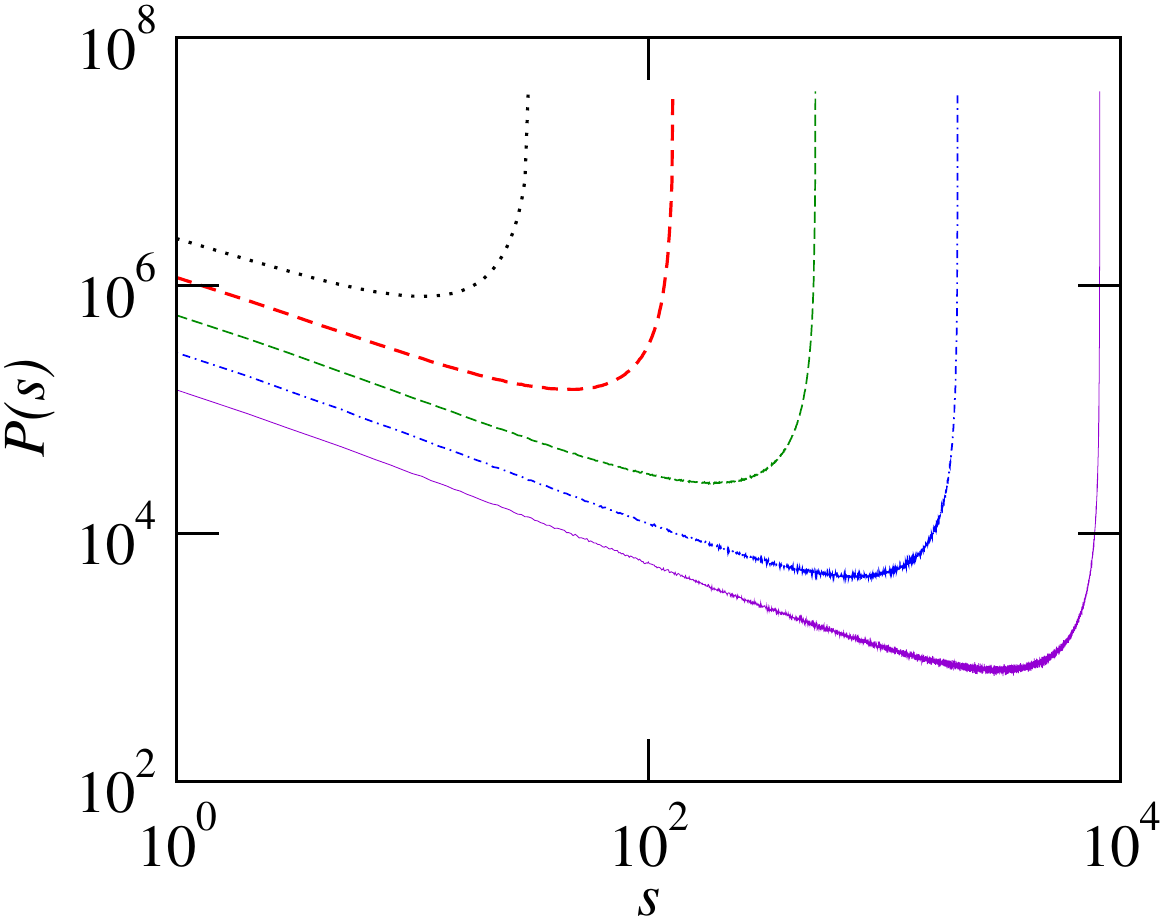}}
	\scalebox{0.68}{\includegraphics{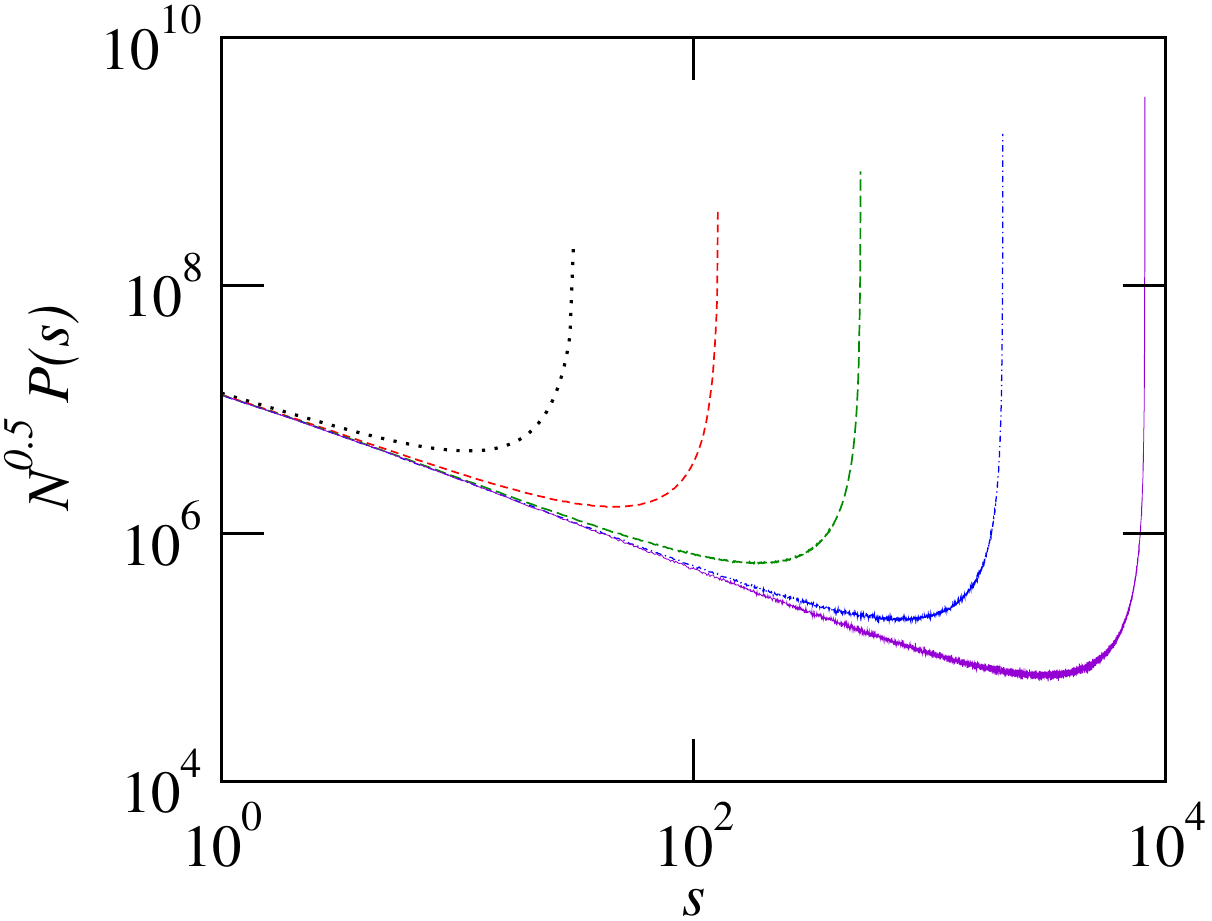}}
	\caption{Left panel: The PDF for $p = f_1(o) = 1/o$ in the moderate input case. The system size varies from $2^5, 2^7, \cdots, 2^{13}$. Right panel: Plot of $\sqrt{N}P(s)$ for different $N$.}
	\label{fig4} 
\end{figure*}

\begin{figure}[t]
	\centering
	\scalebox{0.68}{\includegraphics{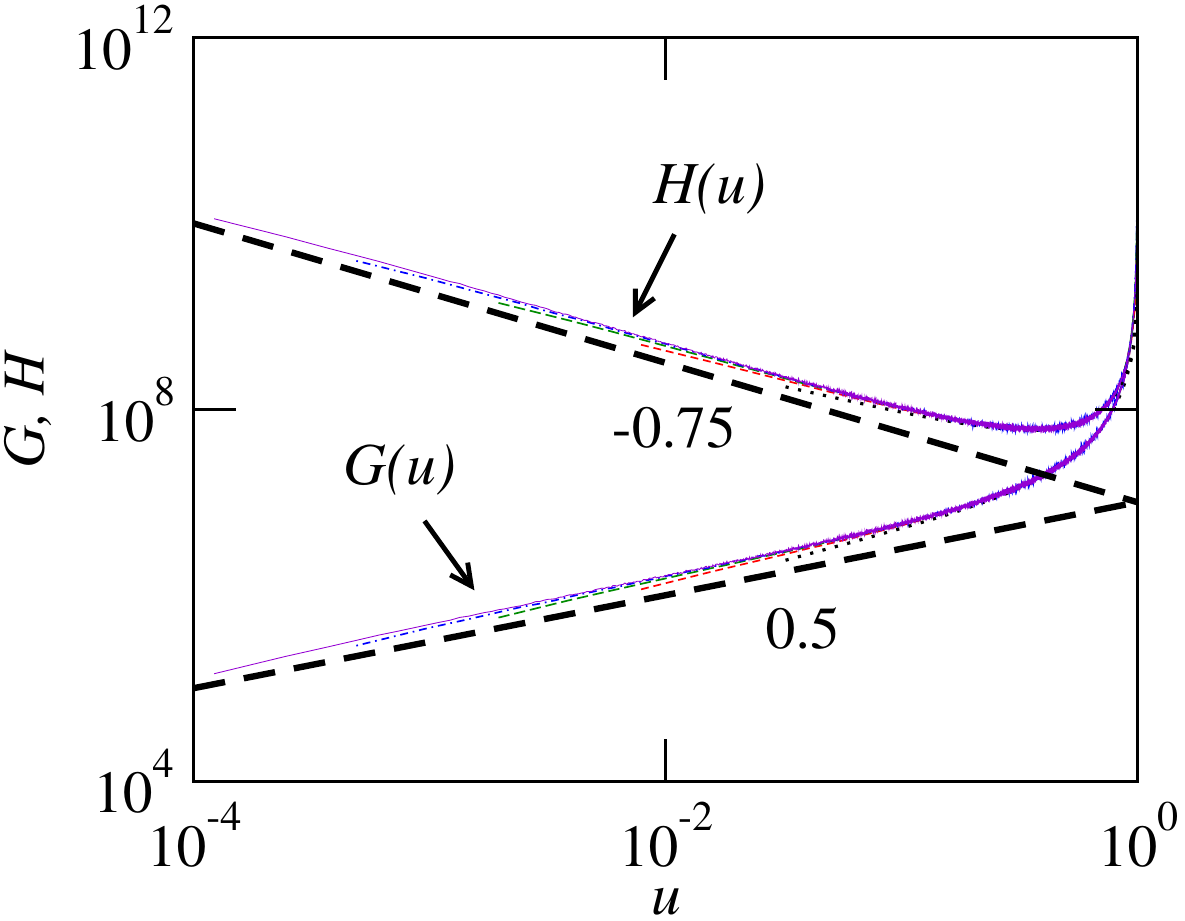}}
	\caption{ The data collapse curves for Fig.~\ref{fig4}~left panel, with $\tau_s+\theta = 1.25$.}
	\label{fig5} 
\end{figure}

\begin{figure}[t]
	\centering
	\scalebox{0.69}{\includegraphics{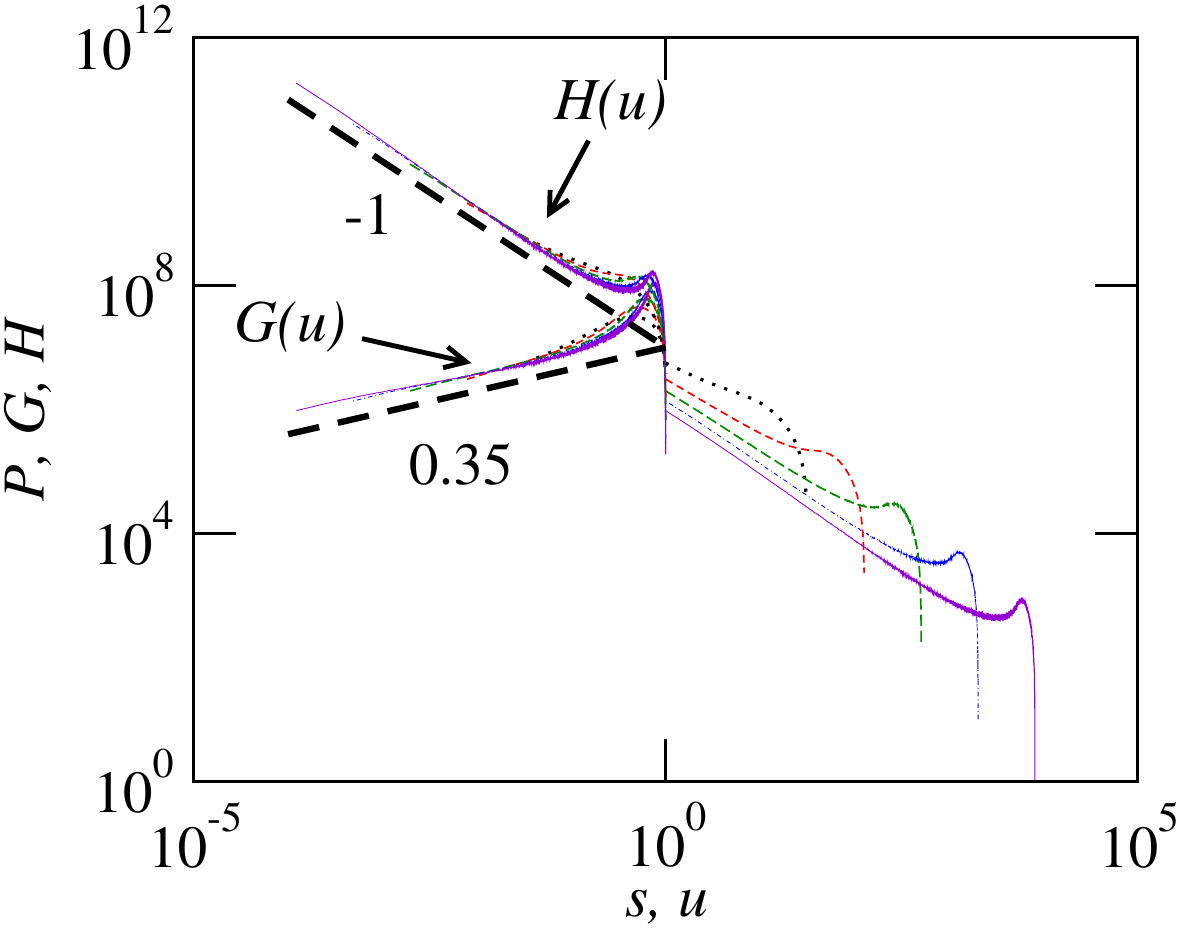}}
	\caption{ The PDF and data collapse curves (with $\tau_s+\theta = 1.35$) for a case where the system is driven in the moderate drive regime with $p = f_2(o) = (o/N)^2$.}
	\label{fig6} 
\end{figure}

\section{Results}\label{sec_4}
In this section, we present our analytical and numerical results obtained in two  drive regimes, namely, no external input and moderate input.  

\subsection{No external input regime}
In the no external input regime, the avalanche size satisfies Abelian distribution \cite{Levina_2014}
\begin{equation}
P_{\alpha,N}(s) = C_{\alpha,N} {{N}\choose{s}}\left[s\frac{\alpha}{N}\right]^{s-1}\left[1-s\frac{\alpha}{N}\right]^{N-s-1},
\label{abel_dist}
\end{equation}   
where the normalization constant is 
\begin{equation}
C_{\alpha,N} = \frac{1-\alpha}{N(1-\alpha)+\alpha}=\frac{1}{N +\frac{\alpha}{1-\alpha}},
\end{equation}
with a parameter $\alpha \in (0, 1)$. Using Stirling approximation $n! \approx \sqrt{2\pi n} (n/e)^n$, Eq.~(\ref{abel_dist}) reduces to
\begin{equation}
P_{\alpha,N}(s) \approx C_{\alpha, N}\frac{1}{\sqrt{2\pi}} \frac{1}{[u(1-u)]^{3/2}}\left[ \frac{1-\alpha u}{1-u}\right]^{N-s-1} \frac{\alpha^{s-1}}{\sqrt{N}},\nonumber
\end{equation} 
where $u = s/N$.

For the critical behavior, the limiting value of the parameters are  $\alpha \to 1$ and $N \to $ large. Applying the limit, we get $\lim_{\alpha\to 1}C_{\alpha, N} \sim 1/N$, and 
\begin{equation}
P(s,N) = \lim_{\alpha\to 1} P_{\alpha,N}(s) \approx \frac{1}{\sqrt{2\pi}} \frac{1}{N^{3/2}} \frac{1}{[u(1-u)]^{3/2}}.
\end{equation}
From Eq.~(\ref{abel_dist}), we note that 
\begin{equation}
P(s=N)  = \lim_{\alpha\to 1} P_{\alpha,N}(s=N) \sim 1/N.
\end{equation} 
 Thus, for the avalanche size $x=s$ with $x_c=N$, the probability distribution is easy to express as
\begin{equation}
P(x, x_c) \sim \begin{cases} x^{-3/2}(1-x/x_c)^{-3/2},~{\rm for~} x< x_c,\\ 1/x_c,~~~~~~~~~~~~~~~~~~~~~~{\rm for ~} x = x_c. \end{cases}
\label{prob_lm}
\end{equation}  
Using Eqs.~(\ref{prob_2}) and (\ref{prob_lm}), the scaling functions can be easily deduced
\begin{equation}
G(u) \sim \begin{cases} (1-u)^{-3/2},~~~~~~~~{\rm for~~} u< 1,\\ \sqrt{x_c},~~~~~~~~~~~~~~~~~~{\rm for ~~} u = 1. \end{cases}
\label{eq_usc_g}
\end{equation}
and
\begin{equation}
H(u) \sim \begin{cases} [u(1-u)]^{-3/2},~~~~~~~~{\rm for~~} u< 1,\\ \sqrt{x_c},~~~~~~~~~~~~~~~~~~~~~{\rm for ~~} u = 1. \end{cases}
\label{eq_usc_f}
\end{equation}

Clearly, $\theta = 0, \tau_s = 3/2$, and $D_s =1$. For $x=t\sim m$, we get $\theta = 0, \tau_t \approx 2$, and $D_t \approx1/2$. We can relate the observable $x$ to $y$ via a simple power-law. And, the scaling theory provides a relation among the critical exponents~\cite{Manchanda_2013}. The model seems to belong to the universality class of the mean-field critical branching process (MFBP). 
Figure~\ref{fig1} shows detailed numerical results for avalanche characteristics, consistent with analytical results [cf. Eqs~(\ref{prob_lm}) to (\ref{eq_usc_f})]. In simulations, we fix the total number avalanches $m = 10^8$ so that we can see a precise dependence of the system size.

\subsection{Moderate input regime}
Recall the dynamics of the model under no external input regime.
Starting with a stable configuration such that $1 \le E_i < M$, randomly pick a node $i$ and activate that by setting $E_i = M$. The active node fires by going into a refractory state $E_i \to 1$, and the connected neighbours get updated as $E_j \to E_j +1$. If neighbour node(s) get(s) activated, the node(s) fire(s) simultaneously. The update occurs in parallel. It eventually leads to the formation of an avalanche event. Note that each node can fire only once during an avalanche event.

In the moderate input regime, an avalanche event of size $o$ is firstly generated as done in the no external input regime. Next, among $N-o$ silent nodes (that did not fire during the initial event of size $o$), we activate $r$ nodes in a random manner (discussed below). In turn, a secondary cascade event arises, and this eventually results in a secondary avalanche of size $s$. When the avalanche event is over, generation of a new independent avalanche event takes place following the same dynamics.

The activation of $r$ nodes takes place with probability $p$. We consider the activation probability $p$, a function of initial avalanche size $o$, as $p = f(o)$. In the simplest case, we assume the probability $p$ to be proportional to $o$, that is, $p = f_0(o) = o/N$.

We note that if $p = f_0$, then the probability distribution for the total size $s+o$ and the secondary size $s$ both $P(s+o)$ and $P(s)$ follow decaying power-law with the same critical exponent $\tau_s \approx 1.15$ [cf. Figs.~\ref{fig2}~and~\ref{fig3}] that is quite close to an exact value of $1.25$~\cite{Das _2019}  within statistical error.     
It suggests that the dominating contribution to critical avalanches is because of the secondary avalanche event.

To understand the extent of scaling features, for the secondary avalanche size probability distribution, we examine cases where $p$ is a nonlinear function of $o$. In general, one can also assume $p$ to be a nonlinear function of $o$. We examined several choices. Here we discuss two representative mathematical examples: (i) $p = f_1(o) = 1/o$ and (ii) $p = f_2(o) = (o/N)^2$. Interestingly, we observe that $P(s)$ still shows scaling but with an explicit system size dependence. This means $P(s)$ is an explicit function of $N$. We find that the probability $P(s)$ decreases on increasing $N$, at fixed $s$, algebraically $N^{-\theta}$ [cf. Fig.~\ref{fig4}]. Also, $P(s+o)$ may not show a clear power law in the nonlinear cases. While we note a partly relaxation of the separation of time scale in the moderate input regime, the secondary avalanche size distribution shows scaling features with non-universal critical exponents. However, the sum of the two critical exponents is approximately constant and close to $\tau_s+\theta = 1.25\pm 0.10$ [cf. Figs.~\ref{fig5}~and~\ref{fig6}].

It is worthy to note that the normalization condition for power-law distribution with an upper cutoff suggests $\tau_s+\theta = 1$ if $\tau_s<1$ and the distribution shows explicit system size dependence~\cite{Yadav_2022_pre}. Although the distribution here shows explicit system size dependence, the sum of the two exponents is significantly greater than 1. While the system size scaling may be a consequence of inherent nonlinearity associated with activation probability, the sum of the two exponents is not 1, possibly because the secondary avalanche size exerts constraint from the initial avalanche event.

\section{Summary and Discussion}\label{sec_5}
In summary, we have examined the properties of critical neuronal avalanches in a neural levels model. The model has been of interest because it is simple and solvable for the avalanche size distribution. It also serves as a representative framework since the branching neural model also shows similar properties. We mainly focused on the scaling functions associated with the critical avalanches. We emphasize that the ``scaling functions" are quite useful. These not only reveals scaling exponent but can also quantify an explicit system size dependence if that exists.

Our intensive simulation studies reveal several interesting features. (1) For no external input regime, we have shown exact results for the scaling functions and verify them numerically. In addition, we numerically confirm an analytical prediction that the probability for $s=N$ is $P(s=N) \sim 1/N$. (2) In the moderate input regime, we observe the following behaviors: (i) The avalanche size has two components initial and secondary sizes. Interestingly, the secondary event dominates when the activation probability is proportional to the initial avalanche size. (ii) Also, the distribution for $s=N$ varies as $P(s=N) \sim 1/\sqrt{N}$.  (iii) If the activation probability is a nonlinear function of initial avalanche size, the secondary avalanche size distribution still shows scaling features with an explicit system size $N^{-\theta}$, meaning the probability at fixed $s$ decreases on increasing $N$. (iv) Remarkably, our numerical estimates suggest that the sum of two exponents is $\tau_s + \theta = 1.25 \pm 0.10$, and it is close to the known value observed for the case when the activation probability is a linear function of initial avalanche size.

Basically, a distinction between initial and secondary cascade events may not be relevant in experimental data, where the system remains under the influence of external stimulation with moderate strength. Our framework of the scaling functions can easily detect complete characteristics of critical avalanches, including an explicit system size dependence. Our analysis is easy to apply to general systems showing critical avalanches.

\section*{ACKNOWLEDGMENTS}
AQ greatly acknowledges support from Inspire Fellowship (DST/INSPIRE Fellowship/IF180689), under the Department of Science and Technology, Government of India. ACY would like to acknowledge seed grants under IoE by Banaras Hindu University (Seed Grant-II/2022-23/48729) and SERB, DST, Government of India (Grant No. ECR/2017/001702).

\end{document}